\documentclass[a4paper,reprint,superscriptaddress,twocolumns,notitlepage,floatfix]{revtex4-1}
\usepackage[utf8]{inputenc}
\usepackage{bm}
\usepackage{amsmath}
\usepackage{mathrsfs}  
\usepackage{braket}
\usepackage{amssymb} 
\usepackage{csquotes}
\usepackage{xcolor}
\usepackage{graphicx}
\usepackage{siunitx}
\usepackage{pgfplots}
\pgfplotsset{compat=1.5}

\usepackage{hyperref}
\usepackage{bookmark}

\definecolor{tab_blue}{HTML}{1F77B4}
\definecolor{tab_orange}{HTML}{FF7F0E}
\definecolor{tab_green}{HTML}{2CA02C}
\definecolor{tab_red}{HTML}{D62728}
\definecolor{tab_purple}{HTML}{9467BD}
\definecolor{tab_brown}{HTML}{8C564B}
\definecolor{tab_pink}{HTML}{E377C2}
\definecolor{tab_gray}{HTML}{7F7F7F}
\definecolor{tab_olive}{HTML}{BCBD22}
\definecolor{tab_cyan}{HTML}{17BECF}

\newcommand{\epsv}{\bm{{\rm \epsilon}}}
\newcommand{\nv}{\bm{{\rm n}}}
\newcommand{\Av}{\bm{{\rm A}}}
\newcommand{\Vv}{\bm{{\rm V}}}
\newcommand{\Sv}{\bm{{\rm S}}}
\newcommand{\Rv}{\bm{{\rm R}}}
\newcommand{\Uv}{\bm{{\rm U}}}
\newcommand{\Nv}{\bm{{\rm N}}}
\newcommand{\kv}{\bm{{\rm k}}}
\newcommand{\delv}{\bm{{\rm \delta}}}
\newcommand{\psixy}{\psi_\text{xy}}

\newcommand{\trans}{^{\scriptscriptstyle T}}

\hypersetup{%
  breaklinks=true,
  colorlinks=true,
  linkcolor=tab_blue,
  filecolor=tab_blue,
  urlcolor=tab_blue,
  citecolor=tab_blue,
}

\def\ee{\rm e}
\def\ii{\rm i}

\def\epsv{\bm{{\rm \epsilon}}}
\def\nv{\bm{{\rm n}}}
\def\Av{\bm{{\rm A}}}
\def\Vv{\bm{{\rm V}}}
\def\Sv{\bm{{\rm S}}}
\def\tildeSvtwo{\bm{{\rm S}}}
\def\dv{\bm{{\rm d}}}
\def\psiv{\bm{{\rm \psi}}}
\newcommand{\psioe}{\psiv_\text{oe}}

\def\psixy{\psiv_\text{xy}}

\begin{document}

\title{Liquid-crystal-based topological photonics}
 \author{Hamed Abbaszadeh}
 \affiliation{Instituut-Lorentz, Universiteit Leiden, Leiden 2300 RA, The Netherlands}
 
 \author{Michel Fruchart}
 \affiliation{James Franck Institute and Department of Physics, University of Chicago, Chicago, Illinois 60637, USA}
 
 \author{Wim van Saarloos}
 \affiliation{Instituut-Lorentz, Universiteit Leiden, Leiden 2300 RA, The Netherlands}

 \author{Vincenzo Vitelli}
 \affiliation{James Franck Institute and Department of Physics, University of Chicago, Chicago, Illinois 60637, USA}
\date{\today}

\begin{abstract}
Topological photonics harnesses the physics of topological insulators to control the behavior of light. Photonic modes robust against material imperfections are an example of such control. In this work, we propose a soft-matter platform based on nematic liquid crystals that supports photonic topological insulators. The orientation of liquid crystal molecules introduces an extra geometric degree of freedom which in conjunction with suitably designed structural properties, leads to the creation of topologically protected states of light. 
The use of soft building blocks potentially allows for reconfigurable systems that exploit the interplay between light and the soft responsive medium.
\end{abstract}
\maketitle

Topological materials are a class of structured materials that exhibit remarkable features such as the existence of chiral edge states robust against backscattering at their boundaries.
These materials inspired from topological insulators (TI)~\cite{Hassan2010} have proven ubiquitous in physics, including examples in photonics~\cite{Hafezi2013,Khanikaev2013,Rechtsman2013,Lu2014,Fruchart2018,Ozawa2019,Rider2019}, mechanics~\cite{Nash2015,Susstrunk2015,Mousavi2015,Huber2016,Zhang2018,Ma2019}, hydrodynamics~\cite{Souslov2017,Souslov2019,Delplace2017,Tauber2019}, stochastic systems~\cite{Murugan2017} and electrical circuits~\cite{Ningyuan2015,Albert2015,Lee2018}.
The unique properties of topological photonic materials suggest several potential applications~\cite{Lu2014,Ozawa2019,Rider2019} ranging from high-power single-mode lasers~\cite{Harari2018,Bandres2018} to slow light~\cite{Guglielmon2019}.

Liquid crystals are soft matter phases characterized by their orientational order~\cite{deGennes1972}.
As a result of this order, liquid crystals can control the propagation of light in a reconfigurable way, with applications ranging from liquid crystal displays (LCD)~\cite{Kawamoto2002} to adaptive lenses~\cite{Algorri2019}.
In this Letter, we show how liquid crystals can be used as a soft-matter platform to realize the building blocks of topological photonics~\cite{Lu2014,Ozawa2019,Rider2019}.
We develop a strategy purely based on liquid crystals where the orientation of the nematic molecules, described by their director field, is used both to realize waveguiding~\cite{Slussarenko2016,Alberucci2016,Jisha2017} and to build Floquet topological materials~\cite{Rechtsman2013,Maczewsky2017,Bellec2017} by coupling these waveguides. 

We first develop a tight-binding model (coupled-modes description) for the coupled liquid-crystal wave\-guides~\cite{Yariv1973}.
We establish its domain of validity through a careful comparison with direct simulations of Maxwell equations.
Along with a precise analysis of the symmetries in the system, these results allow us to engineer a liquid-crystal realization of two archetypal topological systems: a system with non-trivial winding numbers analogous to the one-dimensional Su-Schrieffer-Heeger (SSH) model~\cite{Su1979,Bellec2017} and a system with non-trivial Chern numbers inspired by the two-dimensional Haldane model~\cite{Haldane1988}.

In the paraxial approximation (i.e., at small angles with the direction of propagation), the evolution of light waves in the paraxial direction $z$ can be mapped to the evolution in time of a quantum wavefunction described by the Schrödinger equation. In this picture, modulations of the waveguide in the paraxial direction correspond to a time-dependent potential~\cite{Longhi2009,Szameit2010}.
In Ref.~\cite{Rechtsman2013} it was demonstrated that periodic modulations induced by helix-shaped waveguides allow one to implement photonic Floquet TIs.
In a curved waveguide, the change in the local direction of propagation leads to geometric phases~\cite{Berry1984,Cohen2019} called Rytov-Vladimirskii-Berry phases~\cite{Rytov1938,Vladimirskii1941}, which are eventually responsible for the existence of the photonic TI in Ref.~\cite{Rechtsman2013}.
In contrast, the geometric phases present in our liquid-crystal system, called Pancharatnam-Berry phases~\cite{Pancharatnam1956,Berry1987}, stem from the change in the local optical axes.
As a consequence, the symmetries of the photonic topological material are entirely controlled by the spatial symmetries of the nematic texture. 

\begin{figure}{h!}
\begin{picture}(50,186)
\put(-108,0){\hbox{\includegraphics[scale=0.42]{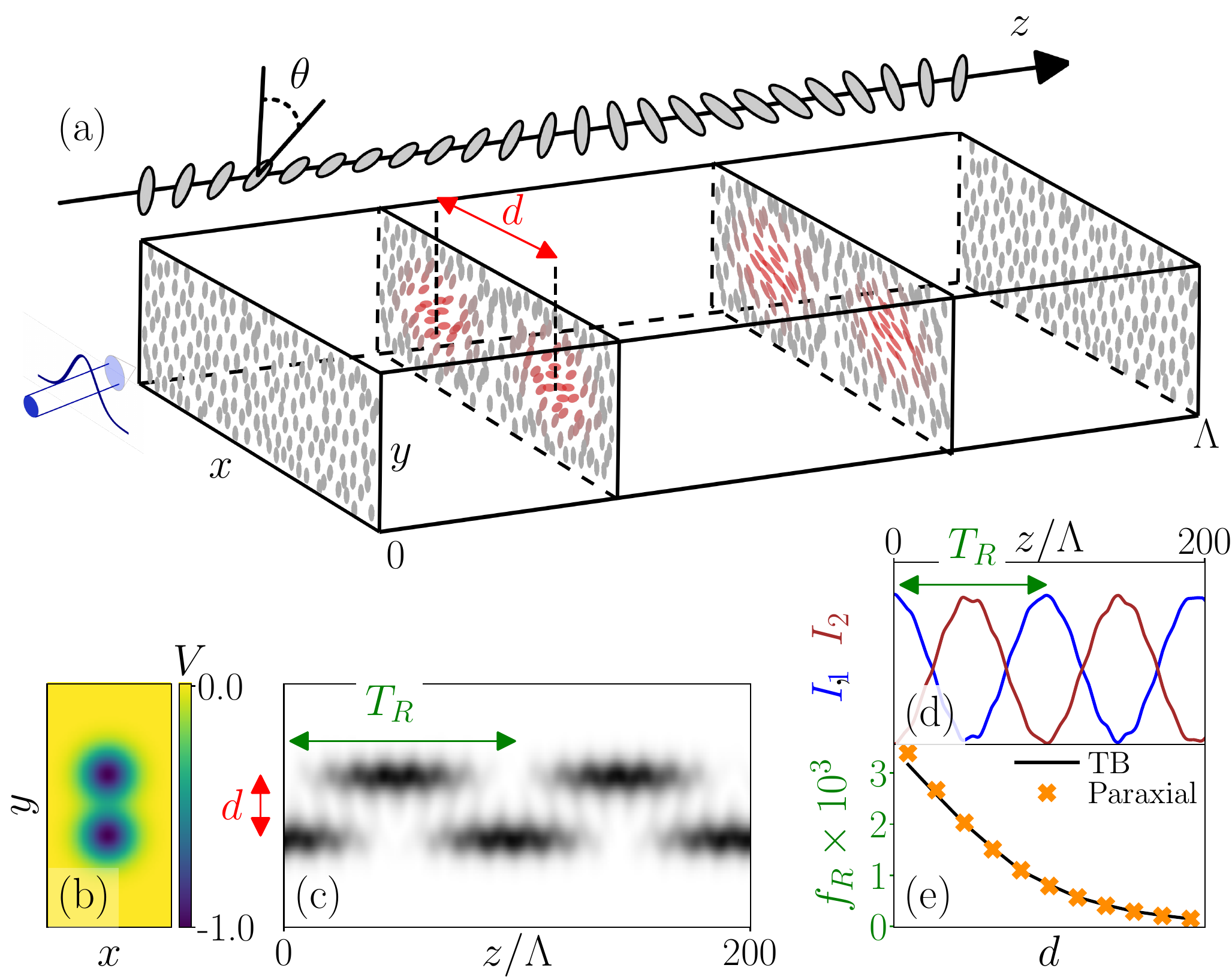}}}
\end{picture}
\caption{Coupling of two waveguides, corresponding to the liquid-crystal pattern in (a) and with an effective optic potential shown in (b). The $z$-axis in (a) shows the orientation of the molecules in the centre of a waveguide for one Floquet period, given by $\Lambda$. (c)-(d) The intensity profile obtained with the guided mode of one of the waveguides as an initial condition shows an oscillatory pattern, which is reminiscent of Rabi oscillations in two-level quantum systems. (e) Dependence of the dimensionless Rabi frequency $f_R = \lambda/T_R$ with the distance between waveguides for both the continuum paraxial simulations and the tight-binding method. The effective interaction between two such waveguides, proportional to $f_{R}$, decays exponentially with their distance.}
\label{fig:Fig2}
\end{figure}

\textit{Light confinement via periodic drive.---}
Consider the propagation of a Gaussian light beam in a uniform nematic liquid crystal, where the nematic rods are in the plane orthogonal to the direction of propagation.
Because of the shape of the nematic molecules, beams with different polarizations experience different refractive indices.
The medium is characterized by two particular indices~$n_\text{o}$ and~$n_\text{e}$, that correspond to the so-called ordinary and extraordinary polarizations that propagate unchanged. 
Hence, three characteristic length scales naturally appear: the light wavelength $\lambda$, the beating length $\Lambda=\lambda/(n_\text{e}-n_\text{o})$ between the ordinary and extraordinary polarizations, and the Rayleigh length $Z_R$ that determines the size of the Gaussian beam.
Additional length scales characterize the spatial pattern of the liquid crystal.
Here, we focus on patterns where only the orientation of the nematic liquid crystal molecules (i.e., the director field) is changed, while the ordinary and extraordinary indices $n_\text{o}$ and $n_\text{e}$ are the same in the whole system.
This orientation is determined by the angle $\theta(x,y,z)$ that the rod-like molecules make with the $x$ axis, as shown in Fig.~\ref{fig:Fig2}(a).
In the regime where $Z_R \gg \Lambda, \lambda$, the dynamics of the electromagnetic field can effectively be described by the evolution of a slowly varying wavepacket $\psi$ which is a two-component vector in the polarization space. In the paraxial approximation, the evolution of this wavepacket is described by~\cite{Longhi2009,Szameit2010,Rechtsman2013,Slussarenko2016}
\begin{equation}
	\label{eq:eom}
	i \, \frac{\partial \psi}{\partial z} = -\frac{1}{2\bar{n}k_0} \left[\nabla_{\perp} + i \Av \right]^2 \psi + \Vv \psi,
\end{equation}
with $\Av(z) = -(\nabla_\perp \theta) {\Sv}(z)$ and $\Vv(z) = -(\partial_z \theta) {\Sv}(z)$, where ${\Sv}(z)$ is a $z$-dependent matrix that only depends on $\Lambda$ and is responsible for the polarization dynamics, see SI.
Formally, this equation resembles the Schrödinger equation of quantum mechanics, provided that the paraxial direction $z$ is replaced by time.

When the orientation $\theta(x,y,z)$ of the director field is periodic in the paraxial direction $z$ with a period $\Lambda$, the Hamiltonian is $\Lambda$-periodic in $z$, and can be analyzed using Floquet theory~\cite{Bukov2015,Goldman2015}.
The main idea is that the propagation of light over large distances $z \gg \Lambda$ is essentially captured by repeating its evolution over one period $\Lambda$, which is described by the evolution operator $U(\Lambda)$ associated with equation \eqref{eq:eom}.
The eigenvalues of the operator $U(\Lambda)$ are phases of the form $\ee^{-\ii \kappa \Lambda}$ where $\kappa$ is the quasi-momentum in the paraxial direction of the corresponding eigenmode. 
Here, the eigenmodes describe guided modes of the soft waveguide~\cite{Slussarenko2016}, and up to variations at the (small) scale of the period $\Lambda$, their intensity remains constant.

\textit{Coupling of waveguides.---}
We now consider a system that consists of two such soft photonic waveguides by transversely repeating the director modulation which corresponds to one waveguide, see Fig.~\ref{fig:Fig2}(a).
When two of these waveguides are located close to each other, they become coupled as light can tunnel from a waveguide to another through evanescent waves: the electromagnetic field inside one waveguide induces a field inside the other one. Fig.~\ref{fig:Fig2}(c) shows that if a guided mode is initially inside one of the two identical waveguides, it will eventually leak into the other one. The light intensity pattern obtained from such interaction oscillates sinusoidally with a period $T_\text{R}$ (see Fig.~\ref{fig:Fig2}(d)) exactly like Rabi oscillations in two-level quantum systems. We confirm these results by directly solving the full Maxwell equations using the finite-difference time-domain (FDTD) method in the open-source software package MEEP~\cite{OSKOOI2010}.

We wish to consider a system made of a large number of coupled waveguides. 
To do so, we need a simplified description of the waveguides and of their couplings, that allows to capture the essential features of the system (such as the Rabi oscillations described above) without having to describe the full liquid crystal configuration. Hence, we use a time-dependent Hückel method~\cite{Huckel1931,Ablowitz2017} to develop a tight-binding model for the photonic waveguides (see SI).
The tight-binding (TB) Hamiltonian $H^{\text{TB}}$ obtained using this method for the evolution of a system of $N$ waveguides reads
\begin{equation}
i\partial_{z}\ket{\psi_n} = \sum_{m=1}^{N} H^{\text{TB}}_{nm}(z) \ket{\psi_m},
\end{equation}
where $\psi_n$ is the mode inside waveguide $n$.
This tight-binding model brings the essential simplicity that is needed to analyze a system with many coupled wave\-guides such as a lattice configuration. 
The Rabi oscillations obtained from this tight-binding model for a two-wave\-guide system are in agreement with the solutions of the Schrödinger equation~\eqref{eq:eom} in the continuum using appropriate initial conditions, validating our approach, see Fig.~\ref{fig:Fig2}(e).
Using this tight-binding model, we further quantify the interaction between two such waveguides and observe that its strength decays exponentially with respect to the distance between them, see Fig.~\ref{fig:Fig2}(e).

\begin{figure}
    \includegraphics[scale=.515]{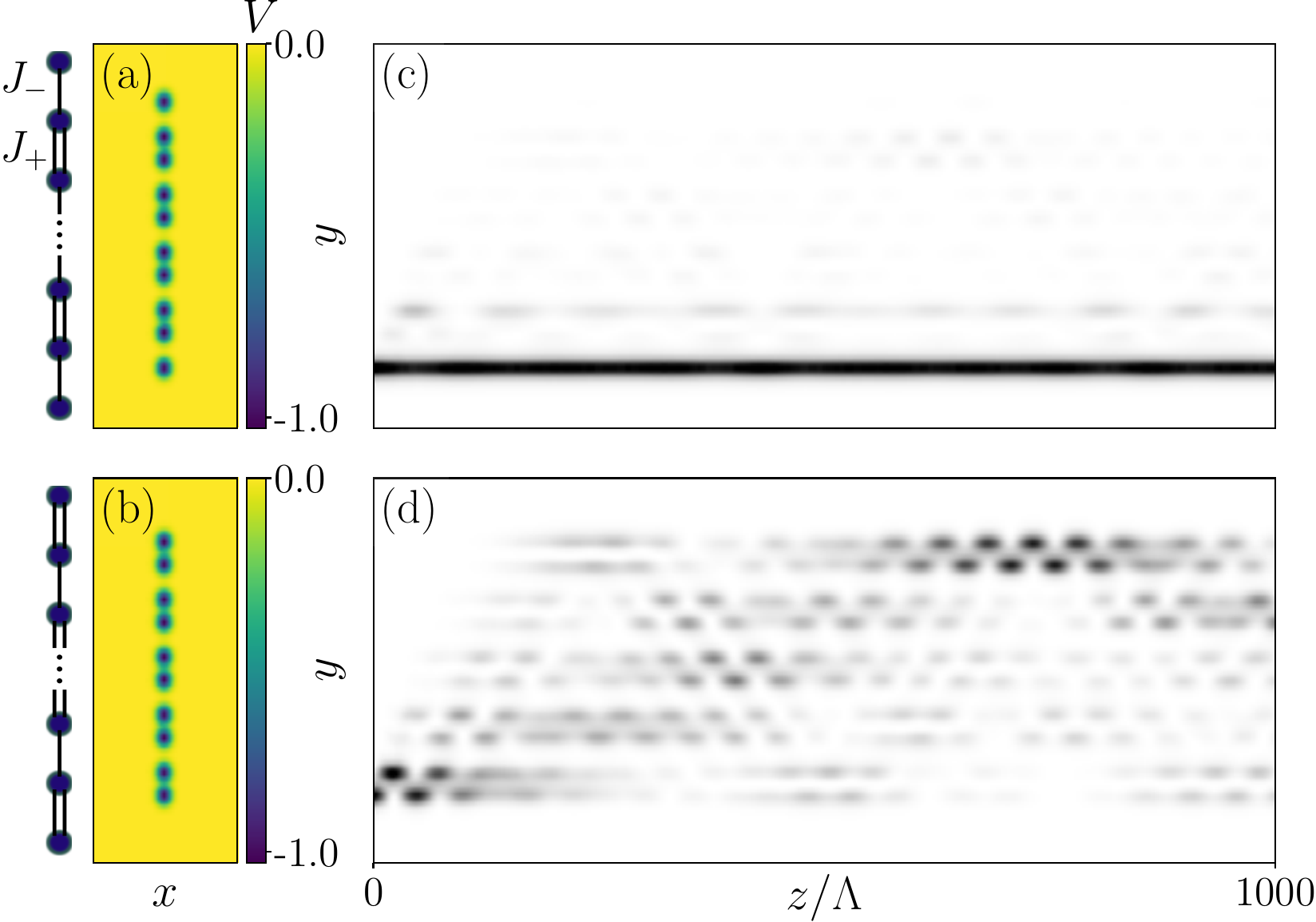}
    \caption{A SSH chain of photonic waveguides in a liquid crystal medium. (a)-(b) The effective photonic potential for two topologically distinct dimerizations of the waveguides in such system which correspond to a SSH chain with and without an edge mode, respectively as shown in (c)-(d). The effective tunneling between these waveguides is controlled by their distance as illustrated in Fig.~\ref{fig:Fig2}(e). Insets in panels (a)-(b) sketch a ssh chain corresponding to each system. Strong ($J_{+}$) and weak ($J_{-}$) couplings are shown by double and single bonds, respectively. (c) The propagation of an edge mode which existence is topologically protected. (d) Shows the scattering into the bulk of the same initial mode in a trivial chain.}
    \label{fig:ssh}
\end{figure}

\textit{Photonic crystals in 1+1d: SSH chain.---}
A lattice of coupled waveguides is obtained by a periodic patterning of the nematic director in the transverse plane. Here, we consider a 1+1d lattice, where the second dimension stands for the paraxial direction $z$ that plays the role of time in this system. 
We consider a system inspired by the Su-Schrieffer-Heeger (SSH) model~\cite{Su1979,Bellec2017}.
Using the interaction-distance dependence from the previous section, we design a lattice of these waveguides on a chain so that the interaction between two neighboring waveguides changes in an alternating way, as shown in Fig.~\ref{fig:ssh}(a-b).
The distance between waveguides is chosen such that the ratio between the two different hopping amplitudes is $J_{-}/J_{+} = \num{0.25(1)}$. 

Figure~\ref{fig:ssh}(a-b) also shows that depending on the ordering of the strong and weak bonds at the boundary, there are two different dimerizations of the neighboring waveguides.
The tight-binding description of the system in Fig~\ref{fig:ssh}(a-b) is a time-dependent version of the SSH chain~\cite{Su1979}. 
We find a photonic state that remains at the edge of one of the two configurations of this system, as shown in Fig.~\ref{fig:ssh}~(c), whereas in the other configuration, in panel (d), the initial mode at the edge leaks into the bulk while it propagates forward. The intensity profile of the localized edge mode shows an exponential decay away from the waveguide at the boundary.
The presence of this edge mode is due to the topology of the Hamiltonian describing the system, which is characterized by an integer winding number across the Brillouin zone (BZ) (see SI).
In this particular case, this topological invariant depends on whether $J_{-}$ is smaller or greater that $J_{+}$, which explains why this edge mode is present in only one of the two configurations in Fig.~\ref{fig:ssh}.

\textit{Symmetries and topological modes in 2+1d.---}
So far, we only considered systems in which there is a symmetry between the photonic modes that propagate forwards and backwards along the $z$ direction. This $z$-reversal symmetry corresponds to time-reversal (TR) symmetry in the effective quantum picture. We would now like to explore the phenomena that can arise with introducing an asymmetry in this direction.
The TR symmetry acts on Eq.~\eqref{eq:eom} through the operator $T = \sigma_z \Theta$, where $\Theta$ is complex conjugation and where the Pauli matrix $\sigma_z$ exchanges right and left circular polarizations.
It follows that a configuration is TR invariant if there is a reference point $z_0$ such that the orientation of the directors satisfies $\theta(z_0-z) = -\theta(z_0+z)$ (see SI).

In the tight-binding model of this system, we focus on the subspace of guided modes, since the unguided ones do not follow a coupled-mode picture. In this reduced description, the TR operator is simply the complex conjugation operator $\Theta$. We prove analytically in the SI that the TR invariance in the paraxial Hamiltonian leads to the TR invariance of the tight-binding model.

\begin{figure*}
    \centering
    \includegraphics[width=\textwidth]{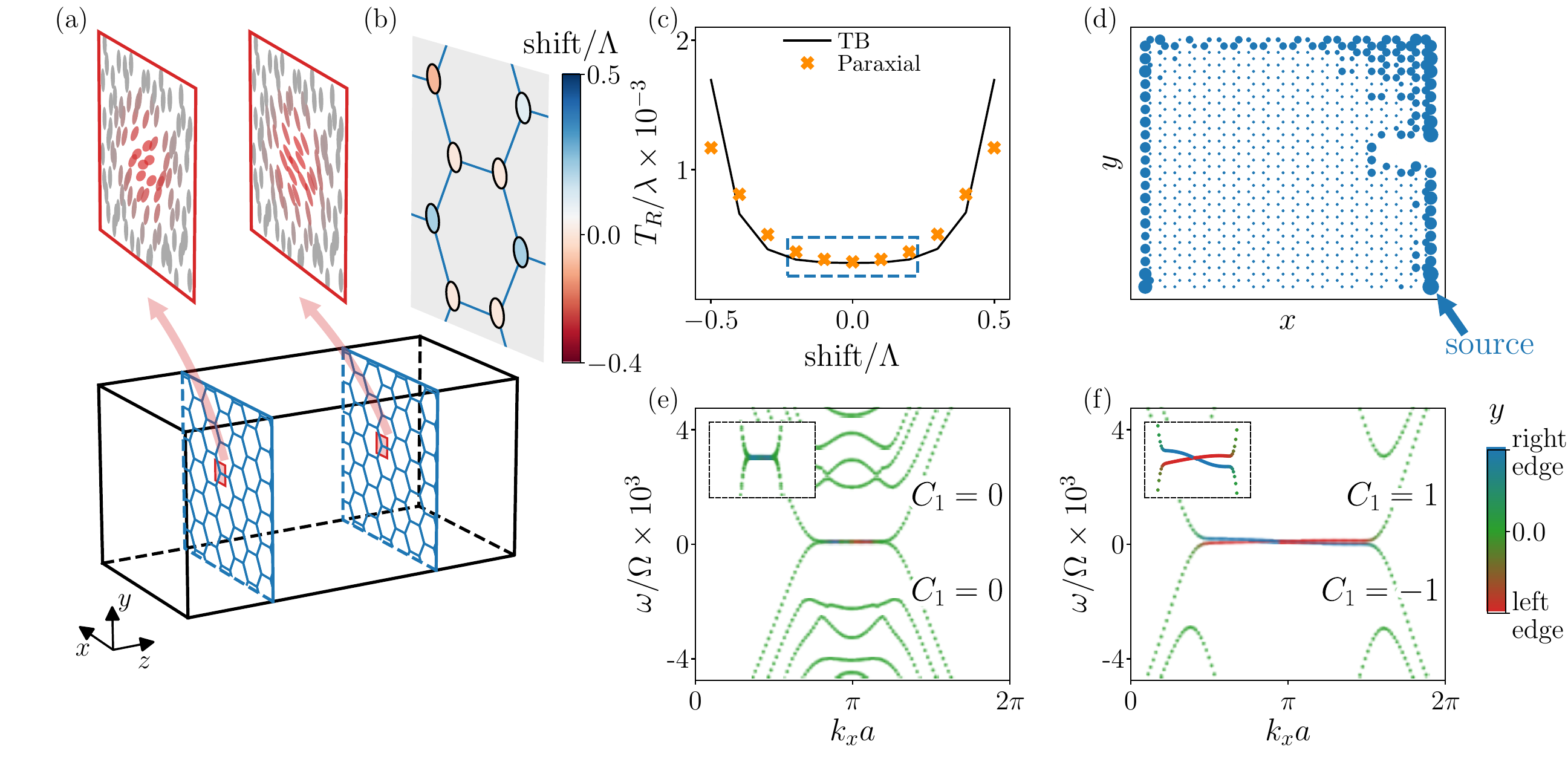}
    \caption{(a) A photonic lattice on a honeycomb structure that is obtained by patterning of the nematic directors in the transverse plane. Each node of the lattice corresponds to a photonic waveguide, as shown in the zoomed-in panels. (b) A unit cell of this photonic lattice, where the waveguides are colored according to their relative phase shift in the $z$-direction. The unit cell is enlarged with respect to the honeycomb one because of the different phase shifts. (c) The Rabi period between two photonic waveguides as a function of the relative shift between them. We focus on the parameter space enclosed by the blue dotted rectangle, where there is a close agreement between the tight-binding model and the continuum paraxial simulations. (d) Evolution of a topological edge mode on the projected $x-y$ plane. The size of each circle corresponds to the light intensity on that site. This mode propagates along the edge without backscattering on obstacles.
    (e)-(f) One sideband in the Floquet band structure of a honeycomb lattice of photonic waveguides (e) without and (f) with relative phase shift. In both cases the TR symmetry is violated using the structural parameter $\eta=0.67$.
    The band structure in (f) shows the presence of one mode at the right (blue) and left (red) edge of this system for a range of transverse momenta in the BZ. The presence of these edge modes and their unidirectional propagation are predicted by the difference between first Chern numbers $C_1$ of the bands which are separated by the gap.}
    \label{fig:2+1d}
\end{figure*}

We break TR symmetry by considering the director field configuration
\begin{equation}
    \theta(x,y,z) = \theta_0(x,y) \left[\sin{\Omega z}+ \eta \, \cos(2\Omega z - \varphi)\right],
\end{equation}
where $\theta_0(x,y)$ describes the nematic pattern in the transverse plane (it is a sum of Gaussians centred at desired positions), $\Omega=2\pi/\Lambda$ is the frequency of the drive (an inverse length scale here), $\eta$ is a dimensionless coefficient that controls the strength of the TR symmetry breaking, and $\varphi$ is the dephasing between the harmonics of the pattern.
We focus on configurations of these waveguides in 2+1 dimensions where the absence of TR invariance can lead to topological modes~\cite{Haldane1988, Rechtsman2013, Ozawa2019}.
A 2+1d lattice of these waveguides can be designed by considering transverse modulations of the nematic directors that are periodic in two directions. We consider a modulation that creates a honeycomb lattice of such waveguides in the transverse plane, as shown in Fig.~\ref{fig:2+1d}(a). 

We find that in our Floquet model, the TR symmetry breaking is not sufficient to get a topological band structure.
This can be understood through a high-frequency Magnus expansion~\cite{Bukov2015} of a general Floquet tight-binding Hamiltonian on a honeycomb lattice, by mapping the obtained effective Hamiltonian with that for the Haldane model~\cite{Haldane1988} (see SI).
We find that breaking the three-fold symmetry between the three neighboring bonds is mandatory to get a non-zero Haldane mass at the first order of the expansion.
Many Floquet driven models involve a rotating gauge field, arising for example from the coupling with a circularly polarized light radiation~\cite{Gomez-Leon2014} or an effective gauge field originating from spin-orbit coupling of light on a helical waveguide~\cite{Rechtsman2013}. In this case, the rotating gauge field effectively breaks the $C_3$ symmetry via a Peierls substitution in the hoppings.

Here, we do not have access to such a rotating gauge field. Instead, we break this symmetry by shifting the waveguides along the z-axis, with a shift that is different for each of the neighboring waveguides, see Fig.~\ref{fig:2+1d}(b). The relative shift of waveguides affects both the strength of their interactions, as shown in Fig.~\ref{fig:2+1d}(c), and induces a dephasing between the hoppings.
We choose a spatially periodic configuration of phase shifts. However, the unit cell is enlarged with respect to the hexagonal lattice, as shown in Fig.~\ref{fig:2+1d}(b) where the colors represent the shifts.

We note that the tight-binding description is only reliable when the shifts are small enough, see Fig.~\ref{fig:2+1d}(c). 
We use the guided mode of each waveguide as a basis for the tight-binding description.
This works well when the waveguides are not shifted. 
When they are, our procedure does not span the whole space where the modes evolve when the waveguides are shifted, because the guided modes of waveguides with relative shifts are different: the guided mode of one waveguide can be repelled from a similar waveguide with a relative shift of origin.
In the following, we focus on the range of shifts where the tight-binding description still provides a reliable approximation (blue dashed rectangle in Fig.~\ref{fig:2+1d}(c)). 

The band structures associated with the tight-binding model of the lattices of waveguides in Fig.~\ref{fig:2+1d}(a-b) on a cylinder are shown in panels (e)-(f). Panel (e) corresponds to a honeycomb lattice of waveguides without relative phase shifts. 
Panel (f) corresponds to the unit cell shown in panel (b), where the waveguides are shifted with respect to each other. In this case, we observe chiral modes localized at the edges of the cylinder.
A direct calculation of the first Chern number of the top and bottom bands (see SI) shows that these edge modes have a topological origin. They circulate unidirectionally along the edge in the transverse plane as they propagate along the $z$-direction despite the presence of a defect at the boundary, see Fig.~\ref{fig:2+1d}(d).
The decay length of the edge modes in the bulk is related to the inverse size of the bulk band gap in which these modes reside, compared to the frequency of the Rabi oscillation. Thus, we find that the decay length is only a few lattice sites, even though the gap is very small with respect to the Floquet frequency.

\textit{Non-Hermitian description of shifted waveguides.---} 
A full description of the photonic crystal should encompass both guided and repelled modes. The analysis above shows that waveguides shifted with respect to each other can be coupled. In this situation, both the guided and repelled modes should be taken into account. In principle, this would entail the use of a non-Hermitian Hamiltonian~\cite{Bender1998} to describe the system, to account for the loss of light intensity due to the repelled light (eventually converted to heat in the bulk of the material). 
We have verified that non-Hermitian effects involving non-orthogonal eigenmodes are negligible in the topological system described in the previous section, validating our approach based on Hermitian topological invariants.
Looking forward, the natural occurrence of non-Hermiticity in the description suggests liquid crystal-based soft waveguides as a promising platform for non-Hermitian optics~\cite{Miri2019,Ruter2010,Feng2012,Feng2017}.

\textit{Experimental considerations.---}
Our work opens up the possibility to harness topological photonics in versatile and technologically relevant liquid crystal platforms.
Our proposal could be realized using commercially available nematic liquid crystals, either with three-dimensional photopatterning techniques~\cite{Sasaki2008}, or by stacking two-dimensional photoaligned slices~\cite{Chigrinov2008,Yaroshchuk2012,
Bisoyi2016,Guo2016,Slussarenko2016}.
For instance, using the same components as in Ref.~\cite{Slussarenko2016} would lead to micrometer-wide waveguides for visible light. 
Further, electrically controlled and light-driven liquid crystals~\cite{Golovin2009,Bisoyi2016} could be exploited to engineer reconfigurable topological photonic devices.

\textit{Conclusion.---}
In this work, we have shown how to realize photonic Floquet topological systems using liquid crystals.
As an example, we have shown how Floquet versions of the SSH and Haldane models can be realized. As photonic crystals, these photonic Floquet TIs are semimetal phases with a strong anisotropy; for instance, a $(2+1)$-dimensional Chern insulator can be seen as a photonic Weyl material~\cite{Noh2017,Lu2015}.
Our analysis based on a reduction of the paraxial wave equation to a tight-binding description provides a blueprint to design photonic structures with targeted topological properties in liquid crystal systems through Pancharatnam-Berry phases. 

\section*{Supplementary Information for: Liquid-Crystal-Based Topological Photonics}
We now provide details for the derivation of Eq.~\ref{eq:eom} in the main text, information for the realization of the model in an actual liquid crystal environment, a derivation of the tight-binding model, and symmetries of the photonic crystal system.

\subsection{Derivation of the paraxial equation of motion}
This section mostly rederives calculations from Slussarenko \textit{\textit{et al.}} ~\cite{Slussarenko2016}.
Electromagnetic wave propagation in a liquid crystal environment can be described by the Helmholtz equation,
\begin{equation}
\label{eq:helmholtz}
\frac{\partial^2}{\partial z^2} \psixy = -\nabla_{\perp}^2 \psixy - k_0^2 \epsv \psixy,
\end{equation}
where $\psixy = (E_x,E_y)$ contains the electric fields for a TE wave, $k_0$ is the light wavenumber in the free space, and $\epsv$ is the dielectric of the medium. In liquid crystals, this dielectric tensor is given in terms of a spatially varying director field $\nv(x,y,z)$ by
\begin{equation}
\epsilon(x,y,z) = \epsilon_\perp \bm{{\mathrm{I}}} + (\epsilon_\parallel-\epsilon_\perp) \nv \trans \nv,
\end{equation}
where $\epsilon_\perp$ and $\epsilon_\parallel$ are the components of the dielectric tensor along the ordinary and extraordinary axes of the molecules, respectively. The refractive indices for these two axes are combined in the following matrix:
\begin{equation}
    \Nv = \sqrt{\epsilon_D} = \text{diag}(n_\text{o},n_\text{e}),
\end{equation}{}
where $\epsv_D = \text{diag}(\epsilon_\perp,\epsilon_\parallel)$ is the diagonalized dielectric tensor.
For a liquid crystal composed of uniaxial elements with orientations in the $x-y$ plane, which is the case that is considered in this paper, the director field is determined in terms of a single parameter, $\theta$, as $\nv = (-\sin{\theta},\cos{\theta},0)$, where $\theta$ is the angle between the extraordinary and the $y$ axes.
One can rewrite Eq.~\ref{eq:helmholtz} for the  ordinary-extraordinary waves using the substitution $ \psiv_\text{oe} = \Rv^{-1}(\theta) \psiv_\text{xy} $, where $\Rv(\theta)$ is the rotation with angle $\theta$, and obtain
\begin{align}
\label{oe}
\partial_z^{2}\psioe &- 2i (\partial_z \theta) \sigma_y\partial_z \psioe = -\nabla_\perp^{2}\psioe  \nonumber\\
&+ 2i (\nabla_\perp \theta) \sigma_y \cdot \nabla_\perp \psioe+ i (\nabla^2 \theta) \sigma_y \psioe \nonumber\\
&+ (\nabla \theta)^2 \psioe - k_0^2 \epsv_D \psioe,
\end{align}
where $\sigma_y$ is the second Pauli matrix.
When the birefringence of the medium, $\Delta n = n_\text{e}-n_\text{o}$ is sufficiently large, we are in a regime where $\Lambda$ is fast compared to the Rayleigh length,
which in terms of the beam's wavelength and its width, $w$ is given by
\begin{equation}{}
Z_\text{R} =  \pi \bar{n} w^2/\lambda,
\end{equation}
where $\bar{n} = (n_\text{e}+n_\text{o})/2$. Thus, we will have two small quantities
\begin{align}
\varepsilon_1 = \lambda/Z_\text{R}
\qquad
\text{and}
\qquad
\varepsilon_2 = \Lambda/Z_\text{R}
\end{align}
in this problem.
In such case, the effect of changes happening over length $Z_\text{R}$ can be considered as perturbations on top of the faster dynamics happening over the birefringence wavelength $\Lambda$. This $z$-dependent interaction can be treated by separating the dynamics over the fast and slow length scales, similar to going to the interaction picture in quantum mechanics. This is done by writing down the wavefunction as $ \psioe(x,y,z,\varepsilon_1z,\varepsilon_2z) = \Uv_\text{oe}(z, \varepsilon_2z) \psi(x,y,z,\varepsilon_1z) $, where
\begin{equation}
    \Uv_\text{oe}(z) = e^{ik_0 \Nv z}
\end{equation}
is the evolution of the polarization in a homogeneous uniaxial medium. The dynamical equation for the wavepacket in the interaction picture can be cast as
\begin{align}
\label{int.pic.}
\partial_z^{2}\psi &+ \left[2i k_0 \Nv -2i (\partial_z \theta) \tildeSvtwo(z) \right] \partial_z \psi = -\nabla_\perp^{2}\psi \nonumber\\
& -2k_0 (\partial_z \theta) \tildeSvtwo(z) \Nv \psi + 2i (\nabla_\perp \theta) \tildeSvtwo(z) \cdot \nabla_\perp \psi\nonumber\\
&+ i (\nabla^2 \theta) \tildeSvtwo(z) \psi + (\nabla \theta)^2 \psi,
\end{align}
where
\begin{subequations}
\label{spinor}
\begin{align}
\tildeSvtwo(z) &= \Uv_\text{oe}^{-1}(z) \sigma_y \Uv_\text{oe}(z) = e^{i\frac{2\pi z}{\Lambda} \sigma_z} \sigma_y \\
&=  \sigma_y \cos(k_0 \Delta n z)  + \sigma_x \sin(k_0 \Delta n z)
\end{align}
\end{subequations}
in terms of the Pauli matrices.
To see how the terms involved in this equation are compared to each other, we write them in terms of dimensionless variables $X=x/w$, $Y=y/w$, and $Z=z/Z_\text{R}$ and then multiply the entire equation by $\dfrac{Z_\text{R}}{2\bar{n}k_0} = \dfrac{w^2}{4}$ to get 
\begin{align}
\frac{\varepsilon_1}{4\pi \bar{n}} \partial_Z^{2}&\psi + \left[i \frac{\Nv}{\bar{n}} - \frac{i}{2\pi\bar{n}} \varepsilon_\theta (\partial_{Z_\theta} \theta) \tildeSvtwo(z/\Lambda) \right] \partial_Z \psi \nonumber\\
=& -\frac{1}{4}\nabla_{\tilde{\perp}}^{2}\psi - (Z_\text{R}/Z_\theta)(\partial_{Z_\theta} \theta) \tildeSvtwo(z/\Lambda) \frac{\Nv}{\bar{n}} \psi \nonumber\\ 
&+ \frac{i}{2} (\nabla_{\tilde{\perp}} \theta) \tildeSvtwo(z/\Lambda) \cdot \nabla_{\tilde{\perp}} \psi \nonumber\\
&+ \frac{i}{4} \left[\nabla_{\tilde{\perp}}^2 \theta + (w/Z_\theta)^2 \partial_{Z_\theta}^2\theta\right] \tildeSvtwo(z/\Lambda) \psi \nonumber\\
&+\frac{i}{4} \left[ \left( \nabla_{\tilde{\perp}} \theta \right)^2 + (w/Z_\theta)^2 \left( \partial_{Z_\theta}\theta\right)^2 \right]  \tildeSvtwo(z/\Lambda) \psi ,
\end{align}
where $\nabla_{\tilde{\perp}} $ is the gradient with respect to dimensionless variables and $Z_\theta$ is a charactristic length scale over which the orientation of the molecules vary along the $z$ axis, and $\varepsilon_\theta = \lambda/Z_\theta$.
Assuming $ \lambda \ll Z_\text{R} $ which implies $ w \ll Z_\text{R} $ will lead to the following dynamical equation after reviving the original variables:
\begin{align}
i \partial_z \psi =& -\frac{1}{2\bar{n}k_0}\nabla_\perp^{2}\psi -(\partial_z \theta) \tildeSvtwo(z) \psi\nonumber\\ 
&  + \frac{i}{\bar{n}k_0} (\nabla_\perp \theta) \tildeSvtwo(z) \cdot \nabla_\perp \psi + \frac{i}{2\bar{n}k_0} (\nabla^2 \theta) \tildeSvtwo(z) \psi \nonumber\\
&+ \frac{1}{2\bar{n}k_0}(\nabla \theta)^2 \psi ,
\end{align}
which can be rearranged into the following form:
\begin{align}
\label{general-eom}
i \partial_z \psi = -\frac{1}{2\bar{n}k_0} &\left[\nabla_\perp -i (\nabla_\perp \theta) \tildeSvtwo(z)\right]^2 \psi -(\partial_z \theta) \tildeSvtwo(z) \psi \nonumber \\ 
&+ \frac{1}{2\bar{n}k_0} \left[ i(\partial_z^2\theta) \tildeSvtwo(z) + (\partial_z\theta)^2\right]  \psi.
\end{align}
The last two terms are of order $ \dfrac{1}{Z_\text{R}}\left( \dfrac{w}{Z_\theta}\right)^2  $, and therefore can be neglected in the limit where $ w \ll Z_\theta $ which will lead us to the Eq.~(1) of the main text.

\subsection{Liquid crystal configuration}
The waveguiding in the system under study is achieved when the following modulation for the director field is used~\cite{Slussarenko2016}:
\begin{equation}
\label{eq:1wvg}
    \theta(x,y,z) = \exp{\left[-\frac{x^2+y^2}{w_D^2} \right]}\sin\left(\frac{2\pi z}{\Lambda}\right),
\end{equation}
where $w_D$ is the width of the Gaussian pattern.
The effective gauge and scalar fields are then determined as
\begin{align}
\Av &= \frac{2(x\hat{x}+y\hat{y})}{w_D^2} \exp{\left[-\frac{x^2+y^2}{w_D^2} \right]} \sin\left(\frac{2\pi z}{\Lambda}\right)\tildeSvtwo(z),\\
\Vv &= - \frac{2\pi}{\Lambda} \exp{\left[-\frac{x^2+y^2}{w_D^2} \right]} \cos\left(\frac{2\pi z}{\Lambda}\right)\tildeSvtwo(z).
\end{align}
Note that the configuration is chosen to be periodic after each beating length. Considering Eq.~\ref{spinor}, it follows that this feature of the liquid crystal structure leads to a $z$-periodic Hamiltonian in the right-hand side of Eq.~\ref{eq:eom}. Therefore, this system can be studied using the machinary of the Floquet Hamiltonians~\cite{Slussarenko2016}.

Note also that the mode evolution equations of this system are linear in $\theta$. Therefore, one can build a lattice of the waveguides above by repeating the modulation of one waveguide, Eq.~\ref{eq:1wvg}, in the transverse plane. For example, a $1$-d SSH chain is described by
\begin{equation}
\theta(x,y,z) = \sum_{i=1}^{N} \theta_i(x,z),
\end{equation}
with
\begin{align}
\theta_i(x,z) = \sin\left(\frac{2\pi z}{\Lambda}\right) \times \\
\biggl ( \exp{\left[-\frac{x-(x_i-d_1/2)^2}{w_D^2}\right]} +\nonumber &\exp{\left[-\frac{x-(x_i+d_1/2)^2}{w_D^2}\right]} \biggr),
\end{align}
where $x_i = x_0 + (i-1) (d_1+d_2) $ are the positions of the centre of pairs of potentials in terms of the alternating distances between the neighboring waveguides, $d_1$ and $d_2$.

\subsection{Floquet tight-binding model}
To build-up a lattice model for a photonic crystal in this system, we exploit the Floquet tight-binding approach. We start by writing down the many-waveguide wavefunction $ \ket{\Phi} $ as a linear combination of single-waveguide modes $\ket{\phi_I}$ as follows:
\begin{equation}
\label{eq:lcao}
\ket{\Phi(z)} = \sum_{I} a_I(z) \ket{\phi_I(z)},
\end{equation}
where $\phi_I(z)$ is obtained by the evolution of the guided mode inside the waveguide $I$. 
The validity of this approximation can be determined by the closeness of the dynamics of this wavefunction to the actual system's evolution. Plugging the wavefunction above into the Schrödinger equation of the system leads to
\begin{align}
H \ket{\Phi(z)} &= i \partial_z \ket{\Phi} \\
&= \sum_J [ i \partial_z (a_J(z)) \ket{\phi_J(z)} \nonumber
 + i a_J(z) \partial_z\ket{\phi_J(z)}].
\end{align}
We can now multiply both sides of this equation with $ \bra{\phi_J} $ to obtain
\begin{align}
&\sum_I i \braket{\phi_J|\phi_I} \partial_z a_I \\
= &\sum_I  \bra{\phi_J} H \ket{\phi_I} a_I\nonumber
- \sum_I \bra{\phi_J} i\partial_z \ket{\phi_I} a_I.
\end{align}
This result can be written as
\def\doubleunderline#1{#1}
\begin{equation}
\label{eq:huckel}
i \partial_z a 
= \doubleunderline{S}^{-1} (\doubleunderline{H}-\doubleunderline{R}) {a},
\end{equation}
where $a = (a_1,a_2,...)$ is a vector in the basis of waveguides, and the matrix elements in this basis are
\begin{align}
\label{eq:tbterm1} \doubleunderline{S}_{JI} &= \braket{\phi_J|\phi_I}, \\
\label{eq:tbterm2} \doubleunderline{H}_{JI} &= \bra{\phi_J} H \ket{\phi_I},\\
\label{eq:tbterm3} \doubleunderline{R}_{JI} &= \bra{\phi_J} i\partial_z \ket{\phi_I}.
\end{align}
Note that these matrices act on the space of guided modes, not on the space of light polarizations.
When starting the coupled-mode theory in Eq.~\ref{eq:lcao}, we project the initial model into the space of the guided modes, which is a one-dimensional subspace of the polarization space.
The effective tight-binding Hamiltonian associated with the continuous problem is then
\begin{equation}
\label{eq:tbhamil}
\doubleunderline{H}_{\text{TB}} = \doubleunderline{S}^{-1} (\doubleunderline{H}-\doubleunderline{R}).
\end{equation}
\subsection{Time-reversal symmetry}
In this section we consider the time-reversal symmetry (TRS) of the effective Schrodinger equation that describes the light propagation in this system. Here we are interested in the behaviour of systems under the inversion of the effective time by the operation $T: z \to -z$. A system is invariant under TRS if there is a $z_0$ such that
\begin{equation}
\label{eq:trs}
T H(z_0+z) T^{-1} = H(z_0-z).
\end{equation}

We first look at the physical set-up that gives a time-reversal invariant system and then derive the tight-binding version of the Eq.~\eqref{eq:trs}. 

\subsubsection{TRS in real space}
The question we would like to answer first is that what features of the liquid crystal system will lead to its invariance under time-reversal. A difficulty in defining a proper time-reversal (TR) operator arises when we notice that the light beam dynamics in this system is extremely affected by its initial polarization. Especial cases of guided and repelled modes for the right/left circularly polarized (RCP/LCP) initial conditions are studied in~\cite{Slussarenko2016}. To avoid potential problems related to this issue, we pick up a TR operator that preserves LCP and RCP lights so that we can study the effect of this operator in the projected space of initial conditions with certain polarization~\footnote{Note that the operator we define does not necessarily work for initial conditions with arbitrary polarization}. This projection can be done by using
\begin{equation}
T = \sigma_z \Theta,
\end{equation} 
where 
$
\sigma_z = \text{diag}(1,-1)
$
is the third Pauli matrix and $\Theta$ is the complex conjugation operator. It then follows that $T \psi_\text{R,L} = \psi_\text{R,L}$. Thus, if the initial polarization of the light beam is a linear combination of the LCP/RCP polarizations (with real coefficients), we have 
\begin{equation}
\label{eq:trsinit}
T \ket{\psi(0)} = \ket{\psi(0)}.
\end{equation}
Now note that the system's wavefunction evolution is given by $\ket{\psi(z)} = U(z) \ket{\psi(0)}$, where
\begin{equation}
U(z) = \lim_{\delta z/z \to 0}\prod_{n=0}^{\left[ z/ \delta z \right]} \exp \left( -i \delta z H(z- n\delta z) \right)
\end{equation}
The time reversal of the wavefunction is
\begin{align}
T &\ket{\psi(z)} \nonumber\\ 
&=T\lim_{\delta z/z \to 0}\prod_{n=0}^{\left[ z/ \delta z \right]}  \exp \left( -i \delta z H(z- n\delta z) \right)  T^{-1} T \ket{\psi(0)} \nonumber \\
&= \lim_{\delta z/z \to 0}\prod_{n=0}^{\left[ z/ \delta z \right]} \left[ T \exp \left( -i \delta z H(z- n\delta z)  \right) T^{-1} \right] T \ket{\psi(0)} \nonumber \\
&= \lim_{\delta z/z \to 0}\prod_{n=0}^{\left[ z/ \delta z \right]} \left[ \exp \left( i \delta z H(-z+n\delta z)  \right) \right] \ket{\psi(0)},
\end{align}
where in the last line we assumed the time reversal invariance of the system around $z=0$ as well as the invariance of the initial wavefunction at the same point given by Eq.~\eqref{eq:trsinit}. The last line is the wavefunction's evolution backward in time till $-z$. Thus, the above calculation actually gives
\begin{equation}
T \ket{\psi(z)} = \ket{\psi(-z)},
\end{equation}
where $\ket{\psi(-z)} = U^{-1}(0,-z) \ket{\psi(0)}$. Apart from its convenience, this relation gives a tool to numerically check whether a given system is TR invariant or not.
    
\begin{figure}
    \includegraphics[scale=.59]{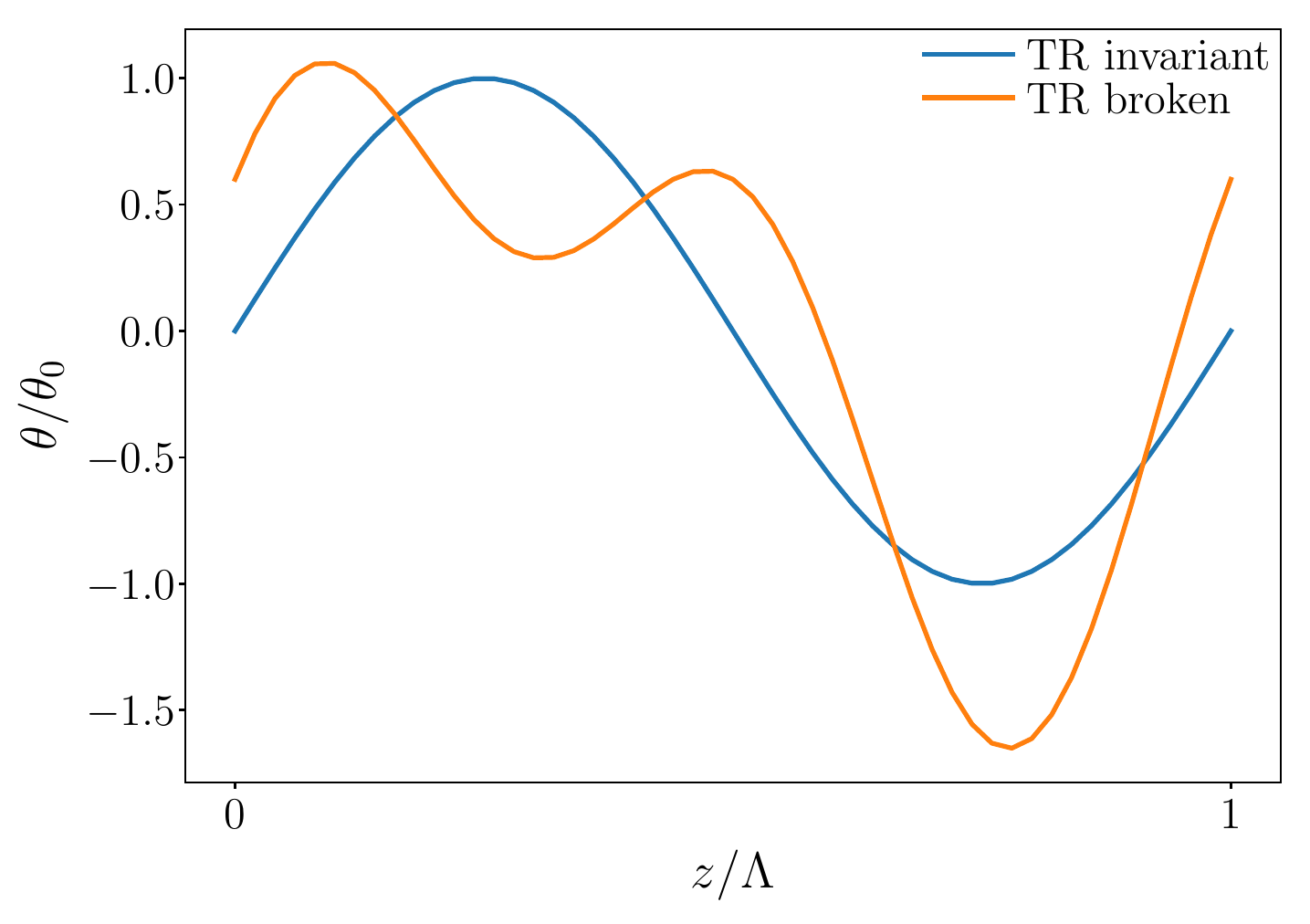}
    \caption{Examples of nematic director fields corresponding to TR invariant and broken systems. The TR broken case is obtained from the Eq.~\ref{eq:trs} in the main text using $\eta=0.67$ and $\varphi=\arctan{0.5}$.}
    \label{fig:TRI}
\end{figure}

We now find the physical systems which are invariant under this TR operator. In other words, we want to find configurations of the system such that the Hamiltonian 
\begin{equation}
H(z) = -\frac{1}{2} \left[\nabla_\perp +i\Av\right(z)]^2 + \Vv(z),
\end{equation}
with $ \Av(z) = -(\nabla_\perp \theta) \tildeSvtwo(z)$ and $ \Vv(z) = -(\partial_z \theta) \tildeSvtwo(z) $ satisfies \eqref{eq:trs}. Now if we use
\begin{align}
T \tildeSvtwo(z) T^{-1} &= T \left[ \cos(2\pi z/\Lambda) \sigma_y + \sin(2\pi z/\Lambda) \sigma_x \right] T^{-1} \nonumber \\
&= \cos(2\pi z/\Lambda) \sigma_y - \sin(2\pi z/\Lambda) \sigma_x\nonumber\\
&= \tildeSvtwo(-z),
\end{align}
we will have
\begin{align}
T H(z) T^{-1} &= -\frac{1}{2} \left[\nabla_\perp -i(-\nabla_\perp \theta)(-z) \tildeSvtwo(-z)\right]^2 \nonumber\\
&+ (-(\partial_z \theta)(-z) \tildeSvtwo(-z)). 
\end{align}
Now one can see that an odd $\theta$ around any point in $z_0$ is a sufficient condition for the system to be TR invariant. Thus the system is TRI when there is a $z_0$ such that
\begin{equation}
\theta(z_0+z) = -\theta(z_0-z)
\end{equation}
We thus can break the effective TRS in this system using the orientation field that is given in Eq.~(3) of the main text. Fig.~\ref{fig:TRI} shows examples of $\theta$ fields which preserve or break the TR symmetry.

\subsubsection{TRS in tight-binding Hamiltonian}
Here, we consider the effect of the time reversal operation in the tight-binding model of the system under study.
It is also insightful to understand the implications of a continuous TRS Hamiltonian on the tight-binding version. Here we consider the terms of $H_\text{TB}$ in Eqs.~\eqref{eq:tbterm1}~-~\eqref{eq:tbterm3} separately. Switching to inner product notation we have for a TR invariant system
\begin{align}
S_{JI}(-z) &= (\psi_J(-z),\psi_I(-z)) \nonumber\\
	       &= (T\psi_J(z),T\psi_I(z)) \nonumber\\
		   &= (\Theta \psi_J(z),\Theta \psi_I(z)) \nonumber\\
		   &= (\overline{\psi_J(z)},\overline{\psi_I(z)}) \nonumber\\
		   &= \overline{(\psi_J(z),\psi_I(z))} = \overline{S_{JI}(z)},
\end{align}
where the bar means the complex conjugation.
\begin{align}
h_{JI}(-z) &= (\psi_J(-z),H(-z)\psi_I(-z)) \nonumber\\
		&= (T\psi_J(z),T H(z) T^{-1} T\psi_I(z)) \nonumber\\
		&= (\Theta \psi_J(z),\Theta H(z) \psi_I(z)) \nonumber\\
		&= (\overline{\psi_J(z)},\overline{H(z) \psi_I(z)}) = \overline{h_{JI}(z)},
\end{align}
and 
\begin{align}
R_{JI}(-z) &= (\psi_J(-z), i \partial_z \psi_I(-z)) \nonumber\\
		&= (T\psi_J(z),i \partial_z T\psi_I(z)) \nonumber\\
		&= (\Theta \psi_J(z), i \partial_z \Theta \psi_I(z)) \nonumber\\
		&= (\overline{\psi_J(z)},-i \partial_z \overline{\psi_I(z)}) \nonumber\\
		&= \overline{(\psi_J(z),i \partial_z \psi_I(z))} = \overline{R_{JI}(z)}.
\end{align}
Thus we conclude that
\begin{equation}
T H(z) T^{-1} = H(-z)
\end{equation}
implies
\begin{equation}
\Theta H_{\text{TB}}(z) \Theta^{-1} = H_{\text{TB}}(-z).
\end{equation}

\subsection{Why is $C_3$-symmetry breaking needed?}
Here we would like to consider the effect of the TRS breaking in a Floquet system. The example of such systems is the Floquet photonic TIs first proposed and observed by Rechtsman \textit{et al.} \cite{Rechtsman2013}. Let us start with a Floquet tight-binding Hamiltonian on a honeycomb lattice. For now, we only consider a nearest neighbour interaction, for which the tight-binding Hamiltonian in the momentum space is given by
\begin{equation}
H(k) = \begin{pmatrix}
M & J(\kv)\\
J^{*}(\kv) & -M
\end{pmatrix},
\end{equation}
where $k$ is the momentum and $M$ is the mass. Let us consider the massless case $M=0$. For a Floquet system, the hopping amplitudes in terms of their Fourier components are given by
\begin{equation}
\label{eq:floq-hopp}
J(\kv) = \sum_{\ell \in \mathbb{Z}} J_{\ell}(\kv) e^{ i \ell \Omega t} =\sum_{\ell \in \mathbb{Z}} \sum_{j=1}^{3} J_{\ell,j} e^{i \kv.\delv_j} e^{ i \ell \Omega t},
\end{equation}
where $\delv_{1,2,3}$ are the three neighbouring vectors, and $\Omega$ is the drive frequency. We drop the explicit $\kv$ dependence notation for now. The Hamiltonian in terms of its harmonics is determined by
\begin{align}
H &= \sum_{\ell \in \mathbb{Z}} H_{\ell} e^{ i \ell \Omega t}, \\
H_\ell &= \begin{pmatrix}
0 & J_\ell\\
J_{-\ell}^{*} & 0
\end{pmatrix}.
\end{align}
We are interested in the large frequency limit of the system that is described by this Hamiltonian. In this limit, the stroboscopic picture gives most of the relevant physics of the problem.
We consider the Floquet-gauge invariant high frequency expansion of the Hamiltonian~\cite{Bukov2015}
\begin{equation}
\label{eq:H_eff}
H_\text{eff} = H_0 + \frac{1}{\Omega} \begin{pmatrix}
M_\text{eff} & 0\\
0 & -M_\text{eff}
\end{pmatrix} + \mathcal{O} \left( 1/\Omega^2 \right),
\end{equation}
where $H_0$ is the time-averaged Hamiltonian and the first-order effective mass is given by
\begin{equation}
M_\text{eff} = \sum_{\ell=1}^{\infty} \frac{1}{\ell} \left[|J_\ell|^2 - |J_{-\ell}|^2 \right].
\end{equation}
We see from here that in the first order of the expansion, the drive behaves as if a mass term is introduced to the Hamiltonian. This effective mass can become non-zero when $J_\ell \neq J_{-\ell}$ for which the TRS needs to be broken.

However, we observe in Fig.~(3)e-f that to have a gapped band structure of this system, the point-group $C_3$ symmetry also needs to be broken. To see why this should be the case, we look back at the Eq.~\eqref{eq:floq-hopp}, where for a $C_3$-symmetric system one can write
\begin{equation}
J_\ell(\kv) =  J_\ell(\kv=0) \sum_{j=1}^{3} e^{i \kv.\delv_j}. 
\end{equation}
This leaves us with an effective mass proportional to the $ \left( \sum_{j=1}^{3} e^{i \kv.\delv_j} \right)^2 $ which is zero and has also a zero first derivative at the Dirac point.
Note that this asymmetry between harmonics of the hopping parameter along different directions is present in the system where a time-periodic rotating gauge field renormalizes the hoppings via Peierls subsitution~\cite{Rechtsman2013}.

Now that we found a way to open up a gap in the Floquet band structure, we still need to investigate if this gap is topological or trivial. To do so, we now take a look at the band structure of a finite system and look for signatures of topological edge modes in their band structure. One important requirement for this gap to be topological is that the mass has different signs on the two inequivalent Dirac points. Let us consider the previous case where we break the $C_3$ symmetry by introducing an asymmetric set of hoppings which satisfy $\dfrac{J_{\ell,1}}{r_{\ell,1}} = \dfrac{J_{\ell,2}}{r_{\ell,2}} = \dfrac{J_{\ell,3}}{r_{\ell,3}} = \kappa_\ell$. Now If we take a look back at the effective mass term in our model, we have
\begin{align*}
&M_\text{eff}(K) = \sum_{\ell=1}^{\infty} \frac{1}{\ell} \left[|\kappa_{\ell}|^2 A_\ell(K)- |\kappa_{-\ell}|^2  A_{-\ell}(K)\right],\\
&M_\text{eff}(-K) = \sum_{\ell=1}^{\infty} \frac{1}{\ell} \left[|\kappa_{\ell}|^2 B_{\ell}(K)- |\kappa_{-\ell}|^2  B_{-\ell}(K)\right],
\end{align*}
where 
\begin{align}
A_\ell(\kv) = B_\ell(-\kv) &= \left| \sum_j r_{\ell,j} e^{i\kv.\delv_j}\right|^2.
\end{align}
Now as we see from here, if we only break the $C_3$ symmetry by a set of real directional hopping scales $r_{\ell,j}$, this will lead to $M_\text{eff}(K) = M_\text{eff}(-K)$, for which the gap in the band structure will be trivial. For a simplified case where the $r$ factors are independent from the harmonic $\ell$, it follows that
\begin{align}
M_\text{eff}(K) &= \sum_{\ell=1}^{\infty} \frac{1}{\ell} \left[|\kappa_{\ell}|^2- |\kappa_{-\ell}|^2 \right] A(K)\\
M_\text{eff}(-K) &= \sum_{\ell=1}^{\infty} \frac{1}{\ell} \left[|\kappa_{\ell}|^2 - |\kappa_{-\ell}|^2 \right] B(K).
\end{align}
For general complex $r$-factors the effective masses can become different.

\subsection{Topological invariants}
Here, we describe the topological indices that we used to characterize the topological features of the physical systems analyzed in the main text. We refer to the reviews~\cite{Hassan2010,Ozawa2019} for more details.

\subsubsection{Topological index for 1+1d}
The dimerized chain's tight-binding Hamiltonian in the momentum space is in general a $2 \times 2$ matrix (due to two sublattices) and can be cast as
\begin{equation}
H(\kv) = \dv (\kv) \cdot \bm{\sigma} + \epsilon \bm{{\mathrm{I}}},
\end{equation}
where $\bm{\sigma} = \begin{pmatrix} \sigma_x,&\sigma_y,&\sigma_z \end{pmatrix}$ is a vector of Pauli matrices, $\dv$ is a vector in the momentum space, and $\epsilon$ is the onsite energy term. When sublattice symmetry is preserved, $d_z=0$ and
\begin{equation}
w = \frac{1}{2\pi i} \int_{\text{BZ}} d^2\kv \ln \dv(\kv)
\end{equation}
counts the number of windings of $\dv$ over the BZ. Since $\dv(\kv)$ is a periodic function, the integration is performed over a closed loop, and thus the parameter $w$ is integer valued and characterizes the topology of the Hamiltonian. For the SSH chain, we find that $w= 1$ for the topological and $w=0$ for the trivial system.
\subsubsection{Topological index for 2+1d}
In this case, we determine the topology of the Floquet-Bloch bands by calculating the first Chern number that is given by~\cite{Thouless1982,Hassan2010}
\begin{equation}
    C_1 = \frac{1}{2\pi i} \int_\text{BZ} d^2 \kv \cdot \bra{n(\kv)} \nabla_{\kv} \ket{n(\kv)},
\end{equation}
where $n(\kv)$ is the eigenmode associated with the $n$th band in each Floquet side band. 
We numerically calculate this index using the method described in Ref.~\cite{Fukui2005}.

 \newcommand{\noop}[1]{}

\end{document}